\documentclass{sig-alternate-10pt}

\usepackage{pslatex}
\usepackage{balance}
\usepackage[margin=1in]{geometry}

\usepackage{pifont}
\newcommand{\xmark}{\ding{55}}%

\usepackage{url}

\newcommand{\spara}[1]{\smallskip\noindent{\bf #1}}

\usepackage{titlesec}

\titlespacing*{\section}
{0pt}{1.5ex plus 1ex minus .2ex}{2ex plus .2ex}
\titlespacing*{\subsection}
{0pt}{1ex plus 1ex minus .2ex}{0.5ex plus .2ex}

\begin{document}

\title{Knowledge Graphs Querying}

\author{Arijit Khan\\
\affaddr{Aalborg University}\\
\email{arijitk@cs.aau.dk}
}

\maketitle

\begin{abstract}
Knowledge graphs (KGs) such as DBpedia, Freebase, YAGO, Wikidata, and NELL
were constructed to store large-scale, real-world facts as
$\langle$subject, predicate, object$\rangle$ triples -- that can also be modeled
as a graph, where a node (a subject or an object) represents an
entity with attributes, and a directed edge (a predicate) is a relationship between two entities.
Querying KGs is critical in web search, question answering (QA),
semantic search, personal assistants, fact checking, and recommendation.
While significant progress has been made on KG
construction and curation, thanks to
deep learning recently we have seen a surge of research on
KG querying and QA. The objectives of our survey are
two-fold. {\bf First}, research on KG querying has been
conducted by several communities, such as databases,
data mining, semantic web, machine learning, information retrieval,
and natural language processing (NLP),
with different focus and terminologies; and also in diverse topics ranging from
graph databases, query languages, join algorithms, graph patterns matching,
to more sophisticated KG embedding and natural language questions (NLQs).
We aim at uniting different interdisciplinary topics and
concepts that have been developed for KG querying.
{\bf Second}, many recent advances on KG and query embedding, multimodal KG,
and KG-QA come from deep learning, IR, NLP, and computer vision domains.
We identify important challenges of KG querying that received less attention
by graph databases, and by the DB community in general,
e.g., incomplete KG, semantic matching, multimodal data, and NLQs.
We conclude by discussing interesting opportunities for the data management
community, for instance, KG as a unified data model and vector-based query processing.
\end{abstract}

\section{Introduction}
\label{sec:introduction}

Knowledge graph (KG) \cite{WeikumDRS21}
is an intelligible data model to support easy integration of data
from multiple heterogeneous sources, providing a formal semantic representation for
inference and machine processing.
One does not require exhaustively modeling their schema; 
new entities and relationships can be added in human-driven, semi-automated, or fully automated
manner to their existing structure without endangering the current functionality.
This follows the semi-structured data paradigm, 
enabling more frequent
and timely updates in knowledge graphs. However, schema-flexibility also introduces challenges
in managing and querying KGs.

\subsection{Challenges in KG Querying}
\label{sec:challenges}

\spara{$\bullet$ Scalable and efficient querying.} The problems are three-fold.
{\bf First}, due to storing cross-domain information and being un-normalized, KGs are massive volume.
Freebase \cite{BollackerEPST08} (Google KG is powered in
part by Freebase)  
alone has over 22 million entities and 350 million relationships in about 100 domains.
Graph-of-Things (GoT) \cite{PhuocQQNH16} 
which is a live
knowledge graph system for Internet-of-Things added
roughly more than 10 billion RDF triples per month.
While some works partition the data across multiple tables, e.g., property tables in Jena2 \cite{WilkinsonSKR03} and Oracle \cite{ChongDES05}, vertically partitioned databases in SW-Store \cite{AbadiMMH09}, etc., many databases store them as one giant table (e.g., 
RDF-3X \cite{NeumannW10}),
or a big graph with labels associated with nodes and edges \cite{
Deutsch19,neo4j
}.
{\bf Second}, KG queries (e.g., {\em  ``find the 10 most commonly followed entities by people within a given
user's second-degree network in the LinkedIn economic graph''}) 
are different from classical relational queries  \cite{SahuMSLO20,BonifatiMT20,MhedhbiS22,2018Bonifati}. They are join-heavy queries over many-to-many relations (e.g., `knows', `follows', `friends' relations), involving recursive joins or graph traversals, and resulting in
complex query shapes, e.g., chain, tree, cycle, star, and flower.
Such queries generate large intermediate results \cite{AGM17
} and query optimization is challenging with traditional
binary join plans.
{\bf Third}, exact subgraph pattern matching via subgraph isomorphism is also NP-complete \cite{GareyJ79}.
For a fixed query pattern, 
subgraph isomorphism can be
verified 
by enumerating all potential candidate matches. 
For node-labeled graphs, if the query pattern has $q$ nodes $v_1, v_2, \ldots, v_q$,
and if the number of candidate node matches (from the data graph) for each query node $v_i$ is $|C(v_i)|$
based on node-label matching, then the search space has size
$\Pi_{i=1}^k|C(v_i)|$. This can be large due to massive data graphs, larger query graphs,
and due to less selective query node-labels. Therefore, even with node-labeled graphs, e.g., KGs,
enumerating all potential candidate matches within the search space is expensive.

Scalability and efficiency of graph query processing (including RDF, KG querying) were studied by data management, theory, and systems communities, 
e.g., graph and join query optimization \cite{MhedhbiS22,
ngo18,
NeumannW09,
SarwatEHM13
},
join vs. graph queries \cite{
SakrEH12,SunFSKHX15,SzarnyasPAMPKEB18
}, indexing \cite{HLPY10,AliSYHN22,
KatsarouNT15}, materialized views
 \cite{IbragimovHPZ16,FanWW14}, efficient (exact) subgraph pattern matching \cite{SL20,
 LeeHKL12},
 multi-query optimization \cite{
 PengGZOXZ21,RenW16}, distributed  processing \cite{AbdelazizHKK17,BouhenniYNK21,KaoudiM15
 }, data partitioning \cite{
 PacaciO19}, I/O efficiency \cite{ZhangCTW13}, caching \cite{KhanSK18,
 PapailiouTKK15}, modern hardware \cite{
 JinY0YQP21,ZengZOHZ20
 }, etc.

\spara{$\bullet$ Flexible schema and semantic matching.} In a KG, similar relationships can be stored in diverse ways,
e.g., for the query, {\em``find all software that have been developed by organizations founded
in California''} on the DBPedia knowledge graph \cite{LehmannIJJKMHMK15}, a recently proposed KG querying system, AGQ \cite{WangKXYPZ22} reports that correct
answers conform to one of at least six different schemas.
It is expected to retrieve all semantically correct (i.e., structurally different, yet ‘relevant’) answers for such queries. If users have full knowledge about DBpedia, they can construct various query patterns or write different SPARQL queries that cover all possible schemas, to obtain all software of interest.
It is challenging for ordinary users to have full knowledge of the vocabulary used in a KG and
the underlying schemas defined in the KG,
since the schema can be large and complex due to heterogeneity,
thus KG querying is difficult.

Additionally, the notion of `relevant' or `correct' answers could very well depend on the user's query intent, or can even be vague,
thus a predefined, `one-size-fits-all' similarity metric might not work in all scenarios.
Data management, semantic web, and ML communities investigated this problem in the context of schema mapping \cite{
RahmB01},
ontology and logic based approaches
\cite{
PoggiLCGLR08,MaLWCP19},
query reformulation \cite{
YaoCHH12,ZouHWYHZ14}, schema-free query interfaces and search \cite{TermehchyWCG12}, approximate subgraph pattern matching \cite{
KhanLYGCT11,KhanWAY13,YangWSY14,ZhengZPYSZ16}, graph simulation, homomorphism, and regular expression based pattern matching \cite{FanLMWW10,MaCFHW14,FanLMTW11}, and KG embedding based query processing \cite{LiGC20,WangKXJHF22,WangKXYPZ22,WKWJY20,HuangZLL19,HamiltonBZJL18,ChenHS22}.   %

\spara{$\bullet$ Incomplete KGs.} Knowledge graphs are incomplete and
follow the open-world assumption --- information available in a KG
only captures a subset of reality.
To retrieve the complete set of correct answers for a given query, one must infer missing edges and relations.
Incompleteness for RDF and property graph data models
received fewer attention \cite{DarariPN14,NikolaouK16}. Dealing with missing graph
structure is more difficult than that with missing attributes on nodes and edges. Researchers studied uncertain graph data management \cite{2018Khan2},  probabilistic knowledge bases \cite{BorgwardtCL18}, and commonsense KGs \cite{IlievskiSZ21}. 
Recent ML approaches
embed a KG and logical queries into the same vector space, to deal with missing edges in the KG \cite{HuangZLL19,HamiltonBZJL18,ChenHS22,RenL20,RenHL20,ZhangWCJW21,ArakelyanDMC21}.

\spara{$\bullet$ User-friendly querying.}
Non-professional users find it difficult to formulate an appropriate graph query, e.g., via SPARQL or subgraph pattern  \cite{BhowmickCL17}, thus more user-friendly approaches were developed:
{\bf (1)} (declarative) graph query languages \cite{AnglesABHRV17},
{\bf (2)}
keyword search \cite{
YangYZ21}, {\bf (3)} query-by-example \cite{MottinLVP14,JayaramKLYE15},
{\bf (4)} faceted search \cite{TzitzikasMP17}, {\bf (5)} visual
query \cite{BhowmickC22,HanFJ11}, {\bf (6)} natural language questions \cite{RoyA20,ChakrabortyLMTL21},
{\bf (7)} incorporating users' feedback \cite{BonifatiCL15,SuYSSKVY15
},
{\bf (8)} query auto-completion and recommendation \cite{LissandriniMPV20}, {\bf (9)} answers explanation \cite{SongNLW22,VasilyevaTBL16,ILL15},
{\bf (10)} conversational QA \cite{ZaibZSMZ22}, etc.  A one-time answer might not be satisfactory.
Exploration-based, interactive methods such as faceted search, users' feedback, query suggestion and completion,
answers explanation, conversational QA were designed, enabling users to refine their queries and obtaining personalized results.

\spara{$\bullet$ Multimodal data querying.} Data are multimodal, consisting of texts, images, and other
multimedia data.
Entities as well as features of both entities and relations
in a KG can have varieties of data types. However, bulk of
KG querying methods only focus on the structured information in triple facts,
since multimodal information are either omitted completely, or are treated as regular
nodes and edges. Thus, KG queries and answers lose richer and potentially useful
information, reducing their effectiveness in downstream tasks.
Recently, multimodal KGs and their querying techniques are 
an emerging area of research
\cite{LiuZ0WJD020,GeseseBAS21}.

\subsection{Related Work and Benefits \\ of Our Survey}
\label{sec:benefits}

The closest to our work are surveys on heterogeneous information networks \cite{ShiLZSY17} and querying attributed graphs \cite{WangLFYC21}. While having similarities, knowledge graphs are complex, modeling real-world facts as $\langle$subject, predicate, object$\rangle$
triples. Different from those surveys, we discuss diverse querying methods on KGs, neural approaches, and graph databases support to process them.

There are surveys on RDF data management and querying \cite{AbdelazizHKK17, AliSYHN22, KaoudiM15, SagiLPH22}, as well as on knowledge graphs \cite{WeikumDRS21,NoyGJNPT19,Abu-Salih21} and its various operations separately, such as KG embedding \cite{AliBHVGSFTL22}, KG reasoning with logics and embedding \cite{abs-2202-07412}, KG-QA \cite{QuamarELO22,ChakrabortyLMTL21}, conversational KG-QA \cite{ZaibZSMZ22}, etc. Surveys on graph databases \cite{LissandriniBV18,T22,BPGFPBAH19,AnglesABHRV17}, queries \cite{2018Bonifati} and optimization \cite{MhedhbiS22}, exact subgraph pattern matching \cite{LeeHKL12} exist.  We mention these important surveys in our article.

Additionally, our contributions are as follows.

-- We unite interdisciplinary topics about KG querying with a taxonomy on KG data models, query classification, databases, querying techniques, algorithms, and benchmarks.

-- We discuss recent neural methods for KG query processing such as KG embedding-based query answering, multi-modal KG embedding, KG-QA, and
conversational KG-QA.

-- We analyze the top-10 commercial graph databases support for KG querying, particularly focusing on
 query languages, user-friendly
and interactive interfaces, KG embedding, and multi-modal KG storage.

-- We emphasize the current challenges and highlight some future research directions.

\subsection{Roadmap}
\label{sec:roadmap}

We stated challenges of KG querying and related surveys in \S \ref{sec:challenges}
and \S \ref{sec:benefits}, respectively. Taxonomy of KG querying with an emphasis on
data models, query classification, languages, technologies, and benchmarks
are introduced in \S \ref{sec:taxonomy}. We highlight deep learning approaches
for KG query processing and QA in \S \ref{sec:neural}. We analyze current graph databases support
for KG query in \S \ref{sec:query_databases} and discuss future directions in \S \ref{sec:future}.

\section{Taxonomy of KG Querying}
\label{sec:taxonomy}

While almost all big data companies, e.g., Google, Microsoft, Facebook, Amazon, IBM, eBay have their proprietary knowledge graphs \cite{NoyGJNPT19
}, many public knowledge graphs are also available, e.g., cross-domain KGs
(DBpedia \cite{LehmannIJJKMHMK15}, Wikidata \cite{VK14}, YAGO \cite{SKW07},
Freebase \cite{BollackerEPST08}, NELL \cite{MitchellCHTYBCM18}), KGs for synonyms
and translations in several languages (BabelNet \cite{NavigliP10}, ConceptNet \cite{speer-havasi-2012-representing}),
domain specific KGs \cite{Abu-Salih21} (COVID19 KG \cite{DuLWCY22}, ClaimsKG \cite{TchechmedjievFB19}), among others.
\begin{figure}[tb!]
	\centering
	\vspace{-2mm}
	\includegraphics[scale=0.3]{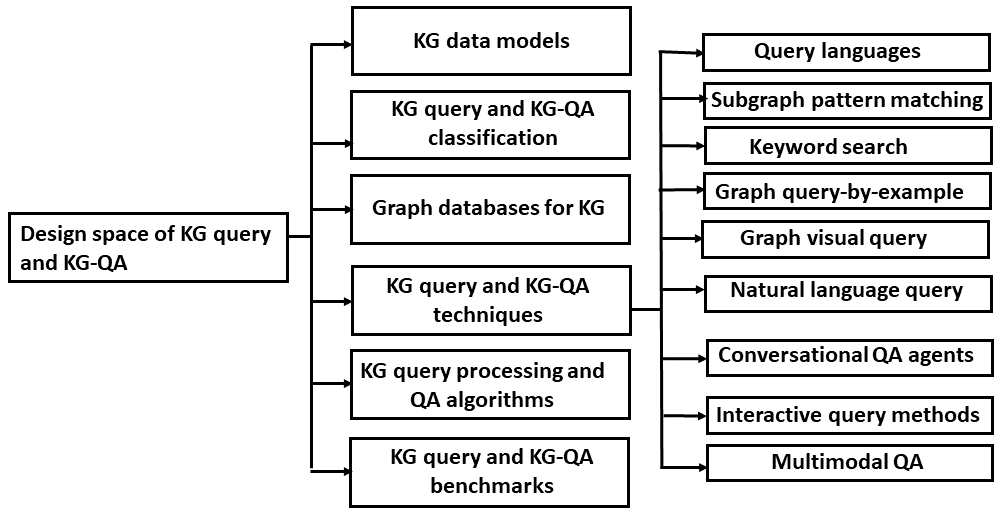}
	\vspace{-5mm}
	\caption{\small {Design space of KG query and KG-QA problems}}
	\label{fig:taxonomy}
	\vspace{-5mm}
\end{figure}

Graph workloads are broadly classified into
two categories \cite{KhanE14}: {\bf (1) online graph queries}
consisting of ad-hoc graph traversal and pattern matching --
exploring a small fraction of the entire graph and requiring
fast response time; {\bf (2) offline graph analytics} with iterative,
batch processing over the entire graph, e.g., PageRank, clustering, community detection, and machine learning algorithms.
Online graph queries and offline graph analytics are also
called {\em graph OLTP} and {\em graph OLAP} (or, {\em graph algorithms}), respectively \cite{T22}.
The focus of this article is read-only online queries without updates in the KG.
KG querying is essential for web search \cite{google}, QA \cite{RoyA20}, semantic search \cite{WKWJY20},
personal assistants \cite{BalogK19}, fact checking \cite{TchechmedjievFB19}, and recommendation \cite{XuRKKA20}.

In this article, we unify various concepts under the broad umbrella of KG query and QA
{with the taxonomy in Figure \ref{fig:taxonomy}, which shows six key design options
for KG query and KG-QA problems: KG data models, KG query and QA classification, graph databases for KG,
KG query and QA techniques, their processing algorithms, and benchmarks.}

\subsection{KG Data Models}
\label{sec:data_model}

Two prominent data models are {\bf (1)} RDF model consisting of $\langle$subject, predicate, object$\rangle$ triples, and {\bf (2)} property graph model having nodes and edges with arbitrary number of properties, where a node (a subject or an object) represents an entity and a directed edge (a predicate) is a relationship between two entities. 
RDF schema (RDFS, also known as an ontology), which is the World Wide Web Consortium (W3C) proposed schema language for RDF, is another RDF (equivalently a directed graph) itself, describing classes, properties, and semantic relationships (e.g., ``is-a'', ``part-of'', ``synonym-to'')
among them. Ontology languages such as OWL have richer vocabulary and define more expressive schema.

\subsection{KG Query and Question Classification}
\label{sec:query_classification}

KG queries and questions can be classified based on several aspects.

\spara{Querying vs. QA.} There are differences between KG query vs. question answering (QA).
A query has a structure, e.g., a graph pattern, a logic query, an SQL or a SPARQL query.
On the other hand, KG-QA {\cite{RoyA20}}
deals with answering unstructured natural language questions (NLQs)
over KGs -- it is a natural language understanding task, that is, semantically parsing an NLQ
to translate it into a query language, such as in SPARQL.

\spara{Simple vs. complex questions.} A simple question involves a single triple and
a single formal query pattern, e.g., {\em ``where was Albert Einstein born?''}
can be answered based on the relation `born': $\langle$Albert Einstein, born, ?place$\rangle$.
On the other hand, a complex question involves multiple KG relations and/or additional
operations, e.g., {\em ``what was the first movie of James Cameron that own an Oscar?''}

\spara{Logic vs. path queries.} First-order logic queries with conjunction, disjunction, negation, and existential quantification
over KGs were widely studied {\cite{ChenHS22}}.
Relational algebra select-project-join (SPJ)
queries and subgraph pattern matching
are conjunctive queries (CQ). A regular path query (RPQ) finds all pairs of nodes connected by at least one path
where the sequence of edge labels on the path follows a given regular expression {\cite{Wood12}}.
A shortest-path query returns the path that has the minimum length between two given nodes {\cite{Sommer14}}.
A conjunctive regular path query (CRPQ) combines CQ (e.g., subgraph pattern matching) with RPQ (e.g., reachability)
{\cite{2018Bonifati}}.

\spara{Factoid vs. aggregate queries.} The answer set to a factoid query is an enumeration of noun phrases, e.g., {\em ``find all movies
by James Cameron”}. An aggregate query retrieves the statistical result of a collection of entities in the answer set, e.g., {\em ``what
is the average length of movies by James Cameron?”} Aggregate queries can be combined with GROUP-BY {\cite{WangKXJHF22}}.

\subsection{KG Query Languages \& Technologies}
\label{sec:query_tech}

A number of technologies exist for KG querying, e.g., SPARQL, SQL extensions, Datalog, graph query languages,
keyword query, exampler query, faceted search, visual query, query templates, natural language questions,
conversational QA, multimodal QA, and interactive methods (e.g., feedback, explanation, suggestion, autocompletion, etc.).

SPARQL is the W3C recommended query language for RDF. 
Microsoft SQL Graph supports SQL extensions that enable creating and querying graph objects \cite{microsoftsqlgraph}.
Graph query languages tend to be declarative like SQL.
Cypher \cite{FrancisGGLLMPRS18}, PGQL \cite{pgql}, and GSQL \cite{DXW18} are declarative graph query languages native to Neo4J, Oracle, and TigerGraph,
respectively. Standardization efforts 
from both academia and industry
led to SQL/PGQ \cite{DeutschFGHLLLMM22}, G-CORE \cite{AnglesABBFGLPPS18}, and GQL  ({\scriptsize\url{https://www.gqlstandards.org/}}). 
Gremlin \cite{Rodriguez15}, adopted by many graph vendors,
is a graph-based programming language that supports both imperative graph traversal and declarative pattern matching.
Datalog-based KG querying was explored in \cite{BellomariniSG18}. Cypher, SPARQL, SQL, and Datalog are not Turing complete.
In contrast, Gremlin and GSQL are Turing complete and hence, more expressive ({\scriptsize{\url{https://info.tigergraph.com/gsql}}}). 

In relational stores, SPARQL queries are reformulated into
SQL queries, then optimized and processed by the relational database management
system. On the other hand, graph-based RDF querying techniques
convert the SPARQL query into a query graph, and perform graph operations (e.g., exact or approximate subgraph pattern matching,
graph traversal) to evaluate the query \cite{
WHCS18}. Recently, \cite{SagiLPH22} investigated which RDF data representations are suitable for what workloads.

More user-friendly means of KG querying involve the following techniques.

\spara{(1) Keyword search} over graphs \cite{
YangYZ21} allows users to provide a list of keywords, and it returns
subtrees/ subgraphs containing those keywords as answers,
based on various ranking criteria, e.g., sum of all edge weights
in the resulting tree/ graph, sum of all path weights from root to each keyword in the tree,
maximum pairwise distance among nodes, etc. While native keyword search algorithms
directly evaluate a keyword query, there are also query reformulation techniques
that convert the keyword query into a more structured format, e.g., SPARQL \cite{ZhouWXWY07} or
query graph \cite{YaoCHH12}. 
Given the set of keywords, the structured queries
are identified by considering term similarity, co-occurrences, and relationships in the KG.

\spara{(2) Graph query-by-example} \cite{MottinLVP14,JayaramKLYE15} enables users to input answer tuple(s) as a query,
and it returns other similar tuples that are present in the knowledge graph.
This follows the well-studied query-by-example paradigm in relational databases,
HTML tables, and entity sets: A user might already know a few answers to the user's
query. The graph query-by-example systems adopt a two-step approach. Given the input
example tuple(s), they first identify the query graph that captures the user's
query intent. Next, they evaluate the query graph to find other relevant answer tuples.

\spara{(3) Faceted search} \cite{TzitzikasMP17} is an explorative, interactive, and
progressive refinement-based search through simple clicks, offering an overview of
the result set at each iteration, thereby assisting
in query formulation according to the 
dataset. Main techniques include faceted taxonomy generation, facet ranking, faceted
interface, visualization, and navigation.

\spara{(4) Graph visual query interfaces} \cite{BhowmickC22,HanFJ11}
allow a user to draw a graph query (e.g., a query graph pattern)
interactively. 
Graph operations such as subgraph matching and enumeration
are employed to evaluate these queries. \cite{HanFJ11}
reformulates graph queries into SPARQL queries.

\spara{(5) Natural language interfaces} \cite{RoyA20,ZouHWYHZ14,
ChakrabortyLMTL21
}
permit users to input questions in natural languages, without requiring them to learn the underlying schema, vocabulary,
or query languages. 
The semantic parsing of a natural language question
involves question analysis, 
phrase mapping and disambiguation,
query construction. Some systems additionally use templates to generate the SPARQL query
\cite{AbujabalRYW18,ZhengYZC18}. Neural approaches are increasingly becoming popular for these tasks.

\spara{(6) Interactive methods} include {\bf (a)} {\em graph query suggestion, expansion, refinement, and autocompletion}
aiming to retrieve more detailed and relevant answers \cite{
LissandriniMPV20}; {\bf (b)} {\em a user's time-bounded search} to
provide `early' answers within the user's response time bound and incrementally improving
the quality of answers with time \cite{WKWJY20};
{\bf (c)} {\em incorporating a user's feedback} for personalized graph querying \cite{BonifatiCL15,SuYSSKVY15
};
{\bf (d)} {\em answer explanation} to support `why', `why-not', `why empty', and `why so many'
questions on query results \cite{SongNLW22,VasilyevaTBL16,ILL15}.

\spara{(7) Conversational QA agents}
\cite{ChristmannRASW19,
ZaibZSMZ22
} engage users in multi-turn QA
to satisfy their information needs.
Once a question is answered, the user can ask another question 
related to the previous QA pair.
Such follow-up questions are usually incomplete, with the context not being clearly specified.
A conversational agent might also ask follow-up questions to understand the user's
query intent. 
Examples include task-oriented systems (scheduling an event),
chat-oriented systems (conducting natural conversations),
QA dialog systems (providing answers about a topic), virtual assistants (e.g., Microsoft Cortana is powered by Microsoft Satori KG),
and {knowledge grounded neural conversation \cite{
ZhouYHZXZ18,TuanCL19,NiPYZC22}}.

\spara{(8) Multimodal QA} \cite{KepuskaB18
} 
consists of multiple user input and output modes
(such as text, image, video, voice, touch, gestures,
gaze, and movements) over multimodal data (including multimodal KGs), having
applications in visual QA, virtual assistants, autonomous vehicles, etc.

A number of keyword search, visual graph query, and natural language query-based interfaces (till 2015) for RDF and KG querying
were compared in \cite{StyperekCSM15,
ElbedweihyWC12} based on their effectiveness and usability.

\noindent $\bullet$ {{\bf Application scenarios of KG querying technologies.}
Writing queries in SPARQL or in other graph query languages requires familiarity with that language,
as well as knowledge of the vocabulary and predicates used in the KG. Such querying modes are
generally suitable for expert programmers and data scientists. Non-expert users and domain scientists
(e.g., biologists, chemists, data journalists, etc. who also use KGs) might prefer more user-friendly
means of asking queries, such as using keywords, graph query-by-example, faceted search, visual
interfaces, and natural language questions. Interactive methods including faceted search,
users' feedback, query suggestion and completion, answers explanation, conversational QA
are helpful in refining users' queries and obtaining personalized results. Conversational and
multimodal QA are critical in virtual assistants.
}

\subsection{Benchmarks for KG Query \& QA}
\label{sec:benchmarks}

Several benchmarks for KG querying and QA exist, such as for simple questions (WebQuestions \cite{BerantCFL13}, SimpleQuestions \cite{BordesUCW15}), complex questions (ComplexQuestions \cite{TalmorB18}, LC-QuAD \cite{trivedi2017lc}), multi-hop questions (HotpotQA \cite{Yang0ZBCSM18}),
conversational QA (ConvQuestions \cite{ChristmannRASW19}), SPARQL query logs \cite{BonifatiMT20},
benchmarks for RDF and SPARQL queries (SP2Bench \cite{SHLP09}, LUBM \cite{GuoPH05}), among others.
QALD is not one benchmark but a series of evaluation campaigns for QA systems over KGs, the recent one being QALD10 ({\scriptsize{\url{https://www.nliwod.org/challenge}}}).

\section{KG Query Processing \& QA: \\ Recent Neural Methods}
\label{sec:neural}

We highlight deep learning advances for KG embedding-based query processing, multi-modal KG
embedding, KG-QA, and conversational QA over KGs.

\subsection{Embedding-based \\ KG Query Processing}
\label{sec:embedding}

KG embedding represents each predicate and entity of a KG as a low-dimensional vector, such that the
original structures and relations in the KG are approximately preserved in these learned vectors \cite{AliBHVGSFTL22}.
{KG embeddings can be broadly classified into four categories.
{\bf (1)} {\em Geometric or translational distance models} compute the plausibility of triples based on a
geometric operation such as a distance function in the embedding space, e.g.,
TransE \cite{BordesUGWY13}, TransH \cite{WangZFC14}, TransD \cite{JiHXL015}, RotatE \cite{SunDNT19}, etc.
{\bf (2)} {\em Semantic matching or tensor decomposition models} compute similarity of latent features by an inner product formulation,
e.g., RESCAL \cite{NickelTK11}, DistMult \cite{YangYHGD14a}, Tucker \cite{BalazevicAH19}.
{\bf (3)} {\em Neural network-based models} generally use convolutional neural networks (CNNs)
to predict the plausibility score of a triple, e.g., ConvE \cite{DettmersMS018}, ConvKB \cite{NguyenNNP18};
or employ graph neural networks (GNNs) which can capture multi-hop relations in the neighborhood of a node,
e.g., RGCN \cite{SchlichtkrullKB18}, CompGCN \cite{VashishthSNT20}, KBAT \cite{NathaniCSK19}, etc.
{\bf (4)} {\em Rule-based models} consider logic rules during embedding learning, e.g., ComplEx-NNE-AER
\cite{WangWGD18} and IterE \cite{ZhangPWCZZBC19}.
}

{For a simple question, if the embeddings of head entity (i.e., head vector $\mathbf{h}$)
and predicate (i.e., predicate vector $\mathbf{r}$) are identified based on the KG embedding, link prediction
can be employed to infer the tail entity, e.g., tail vector $\mathbf{t}\approx\mathbf{h}+\mathbf{r}$ via TransE.
EAQ \cite{LiGC20} applies KG embeddings and uses spatial indexes to efficiently answer top-$k$ and aggregate queries.}

{We categorize recent deep learning techniques for KG query processing into two classes -- both
categories evaluate input graph query patterns and can deal with incomplete KGs and schema mismatch between the query and a KG.}

{{\bf (1)} {\em Query answering methods trained on single-hop queries}, e.g.,
\cite{LiGC20,WKWJY20,WangKXJHF22,ArakelyanDMC21,GZRT22},
though trained on single-hop queries, can process multi-hop and complex input queries by first decomposing complex queries
into smaller subqueries and then combining the answers of subqueries in a systematic way.
For instance, \cite{WKWJY20,WangKXJHF22} process queries having complex shapes (chain, cycle, star, and flower),
aggregate functions (COUNT, SUM, AVG), FILTER and GROUP-BY operators over KG embedding.
Since KG embedding techniques deal with $\langle$subject, predicate, object$\rangle$
triples and are similar to training with single-hop queries, these query answering methods can directly work with
KG embedding, without separate single-hop training queries and their answers.}

{{\bf (2)} {\em Query answering methods trained on multi-hop queries} \cite{SunAB0C20,RenHL20,ZhangWCJW21,LiuDJZT21,YangQLLL22,LongZAWL22,HuangCL22,ChenHS22,Zhu0Z022}
embed multi-hop queries and their answers (i.e., entities from a KG) close to each other in the
same embedding space. These methods deal with logical queries, often implement logical operators in
neural ways, and significantly reduce query processing
time via inference. Unlike generating large intermediate results due to decomposing complex queries
into smaller subqueries, these approaches reduce query answering to dense similarity matching of query and
entity vectors. They can further be classified as geometry, distribution, or fuzzy logic-based methods
according to generated embeddings. The former embeds entities and queries with geometric shapes. Examples
include Query2box \cite{RenHL20}, NewLook \cite{LiuDJZT21}, and ConE \cite{ZhangWCJW21}.
Distribution-based approaches encode entities and queries into probabilistic density, e.g.,
BetaE \cite{RenL20}, GammaE \cite{YangQLLL22}, NMP-QEM \cite{LongZAWL22}.
Fuzzy logic-based methods (e.g., FuzzQE \cite{ChenHS22}, ENeSy \cite{Zhu0Z022})
define logical operators in a learning-free manner following fuzzy logic,
whereas only entity and relation embeddings require learning.
Geometry and distribution-based approaches are trained with complex queries and their answers,
which can be generated by crowdsourcing \cite{BerantCFL13},
or by automatic generation from a KG as in \cite{RenDDCZLS22
}.
Fuzzy logic-based methods can be trained on single-hop or complex queries.
Different from the above approaches, kgTransformer \cite{LiuZSCQZ0DT22}
uses a Transformer-based GNN architecture, models logical queries as masked
prediction, and proposes a masked pre-training strategy.
}

\subsection{Multi-modal KG Embedding}
\label{sec:multimodal}

Multi-modal data (e.g., text, image, multi-media) is associated as attributes
of entities and relations, or treated as new entities in a KG. Multi-modal KG embeddings are critical
for querying multi-modal KGs, and can be classified as follows.

\spara{KG+text.} Notable methods are Extended RESCAL \cite{NickelTK12}, DKRL \cite{XieLJLS16},
and KDCoE \cite{ChenTCSZ18} that embed KGs having textual descriptions of entities. These methods vary in how entity embedding from text is obtained
(e.g., via CNN, LSTM, bag-of-words, etc.) and then how it is combined with structure-based representation.
Recently, efforts were made to combine pre-trained language models with KG+text embedding, {e.g., {\bf (1)} {\em when KGs having textual
description of entities}: SimKGC \cite{WZWL22}, KEPLER \cite{WangGZZLLT21}, KnowlyBERT \cite{KaloFEB20}, K-BERT \cite{LiuZ0WJD020};
{\bf (2)} {\em when KGs and text data are stored separately}: DRAGON \cite{YBRZMLL22}, JAKET \cite{Yu0Y022}, OREO-LM \cite{HuX0WYZCS22}, DRLK \cite{ZhangD0H22}.
}

\spara{KG+image.} IKRL \cite{XieLLS17}, 
RSME \cite{WangWYZCQ21}, and
{MuKEA \cite{DingYLHC022}} learn
KG embedding by jointly training a structure-based representation (e.g., TransE)
with an image-based representation obtained via image encoder. 

\spara{KG+text+image.} To embed KGs having texts and images, several models were proposed,
e.g., {Knowledge-CLIP \cite{PYHSH22}},
CMGNN \cite{FZHWX22}, MKBE \cite{PezeshkpourC018}, MKGAT \cite{SunCZWZZWZ20}, TransAE \cite{WangLLZ19}.
They employ various neural encoders for multi-modal data and combine them with existing relational
models.

\subsection{Neural Methods for KG-QA}
\label{sec:parsing_qa}

Answering natural language questions (NLQ) over knowledge graphs
involve several subtasks including entity linking,
relationships identification, identifying logical and numerical operators,
query forms, intent, and finally the formal query construction \cite{QuamarELO22}.
Rule-based methods use ontologies and KG for phrase mapping and disambiguation
to link entities and relations to the KG, and then employ grammars to generate formal queries.
Recently, neural network-based semantic parsing algorithms have become popular for KG-QA,
which are categorized as classification, ranking, and translation-based \cite{ChakrabortyLMTL21}.
Classification-based parsing algorithms rely on classification models to predict the relation
and entities in a simple NLQ. For more complex questions, ranking-based methods
employ a search procedure to find the top few probable query candidates, followed by
using a neural network-based ranking model to find the best candidate.
Translation based KG-QA methods employ a sequence-to-sequence model, consisting of decoder
and encoder to translate a natural question into a formal query.
Based on the types of training data, their training methods can be fully
supervised (consisting of NLQs and their formal queries during training)
or weakly supervised (provided with NLQs and their execution results, but without their formal
queries during training).

{
More recently, \cite{
HuangZLL19,CaiZWW21,RenDDCYSSLZ21,SaxenaTT20,
LiuDXXT22,SaxenaKG22} propose methods to answer
NLQs over KGs in an end-to-end manner.
They can deal with incomplete KGs, semantic meaning of NLQs, and
ambiguity of entity names and relations.
KEQA \cite{HuangZLL19} jointly learns head entity, predicate, and tail entity
representations of a simple NLQ in a given KG embedding space. Attention-based BiLSTM models are
used for the head entity and predicate representation learning.
EmbedKGQA \cite{SaxenaTT20} learns representation of a multi-hop NLQ in the
KG embedding space first by using RoBERTa (robustly optimized BERT pretraining),
followed by fully connected linear layers with ReLU activation, and finally projecting onto the
KG embedding space. DCRN \cite{CaiZWW21} identifies informative
evidence from candidate entities in a multi-hop question by using their semantic information,
then finds answers by performing RNN encoder-decoder-based sequential reasoning
following the graph structure on the retrieved evidence.
LEGO \cite{RenDDCYSSLZ21} alternates between
growing the query tree and the reasoning action in the KG embedding space.
BiNet \cite{LiuDXXT22} uses an encoder-decoder-based model that transforms
multi-hop NLQs into relation paths, and jointly addresses knowledge graph completion
and KGQA tasks. KGT5 \cite{SaxenaKG22} employs an encoder-decoder Transformer model,
with pretraining the model on the KG using the link prediction task,
and then the model is fine-tuned for complex question answering.
}

\subsection{Conversational QA on KG}
\label{sec:conversational}

Conversational QAs are extensions to one-shot NLQs,
involving a sequence of questions and answers that appear as a dialogue between
the system and the user \cite{RoyA20,QuamarELO22}. 
Conversational QA systems involve
dialog manager and response generator to keep track of the dialog history
and for generating natural language responses, respectively.
Sequence-to-sequence and pre-trained language models are
used for these tasks.

Knowledge grounded neural conversation models
generate more informative responses.
To understand the context of
follow-up questions, commonsense KG-based context expansion is useful \cite{ZhouYHZXZ18}.
DyKgChat \cite{TuanCL19} zero-shot adapts to dynamically updated knowledge
graphs during conversation. HiTKG \cite{NiPYZC22} proposes a hierarchical Transformer-based
graph walker model, which learns both short-term and long-term conversation goals.

\noindent $\bullet$ {{\bf Interaction between neural and classic approaches.}
We identify scenarios where neural and classic KG query processing and KG-QA can be
complementary to each other. {\bf First}, neural semantic parsing translates
NLQs into structured queries, e.g., SPARQL queries or subgraph patterns, and
classic approaches can be applied to evaluate them. Classic approaches identify
intermediate results that help interpreting each step in query processing.
{\bf Second}, neural approaches can also assist in interactive, exploration-based
query processing by automated query suggestion and completion, incorporating
user's feedback, and providing personalized results.
}

\section{Graph Databases Support for KG Query}
\label{sec:query_databases}

We analyze the top-10 commercial graph DBMS according to
{\scriptsize\url{https://db-engines.com/en/ranking/graph+dbms} (accessed on December 30, 2022)},
which ranks commercial database management systems based on their popularity.
In the past, graph databases were benchmarked in regards to their performance,
database systems offerings, data organization techniques, queries, etc. \cite{LissandriniBV18,T22,BPGFPBAH19,AnglesABHRV17}.
Different from them and following our taxonomic discussion, we
categorize which graph DBMS supports what data models, query
languages, user-friendly and interactive interfaces.
Given the popularity of deep learning and KG embedding that are critical for incomplete or multimodal KG querying,
we also investigate if these graph databases support graph embedding and multimodal KG-QA.
Our findings are summarized in Table~\ref{tab:graph_dbms}.

\begin{table*}[t!]
    \centering
    \caption{\small Categorization of top-10 commercial graph DBMS based on KG data models, query languages, user-friendly and interactive
    interfaces, support for graph embedding and multimodal KG-QA. PG: property graph, RDF: RDF triples.}
    \smallskip
    \begin{scriptsize}
    \begin{tabular}{c||c|c|c|c|c}
         {\bf graph DBMS} & {\bf KG data models} & {\bf query languages}    & {\bf user-friendly \& interactive interfaces} & {\bf graph embedding} & {\bf multimodal KG-QA}   \\ \hline \hline
         Neo4J     & PG & Cypher, Gremlin  & Popoto.js: create interactive visual query; & \checkmark  & \xmark \\
                   &     &                 & Neo4j Bloom: write patterns similar to NLQs &             &        \\ \hline
         Microsoft & PG & Gremlin & 3rd-party data visualization tools                   & \xmark & \xmark \\
         Cosmos DB &    &         & (e.g., Graphlytic, Graphistry, Linkurious)           & & \\ \hline
         Virtuoso  & RDF& SPARQL  & faceted browsing, 3rd-party tools (e.g., LodLive) & \xmark & \xmark \\ \hline
         ArangoDB  & PG & AQL, Gremlin & graph viewer, 3rd-party tools (e.g., Cytoscape)& \checkmark & \xmark \\ \hline
         OrientDB  & PG & SQL-like, Gremlin & OrientDB studio: visualize graphs and schema & \xmark & \xmark \\ \hline
         JanusGraph& PG & Gremlin & 3rd-party tools (e.g., Cytoscape) & \xmark & \xmark \\ \hline
         Amazon    & PG,& Gremlin, & Neptune Workbench & \checkmark & \xmark \\
         Neptune   & RDF& SPARQL  & & & \\ \hline
         GraphDB   & RDF& SPARQL & faceted search, 3rd-party tools (e.g., metaphactory) & \xmark & \xmark \\ \hline
         TigerGraph& PG & GSQL    & TigerGraph GraphStudio & \checkmark & \xmark \\ \hline
         FaunaDB     & PG & GraphQL & \xmark & \xmark & \xmark \\ 
    \end{tabular}
    \end{scriptsize}
    \label{tab:graph_dbms}
    \vspace{-3mm}
\end{table*}

\spara{(1) Neo4j} \cite{neo4j} provides a native graph database with property graph data model and
Cypher query language. It also supports the Apache TinkerPop ({\scriptsize{\url{http://tinkerpop.apache.org/}}})
 acting as a connectivity layer to use Gremlin. Neo4J supports several graph analytic tools (e.g., Popoto.js, Neo4j Bloom)
that assist in interactive, visual query building and suggestion. Neo4J's graph data science library
implements three graph embedding methods (FastRP, GraphSAGE, and Node2Vec), node classification and regression, link prediction.

\spara{(2) Microsoft Cosmos DB} for Gremlin (graph) ({\scriptsize{\url{https://learn.microsoft.com/en-us/azure/cosmos-db/gremlin/introduction}}})
is a hybrid graph DB service on top of Microsoft's NoSQL Azure Cosmos DB.
It follows the Apache TinkerPop specification using Gremlin as the query language.
The graph data can be visualized and explored via third-party tools,
e.g., Graphlytic, Graphistry, Linkurious.

\spara{(3) Virtuoso} ({\scriptsize{\url{https://virtuoso.openlinksw.com/}}})
is a hybrid database which stores KGs as RDF triples and provides a SPARQL endpoint.
Besides Virtuoso faceted browsing, third-party tools (e.g., LodLive \cite{CMA12})
exist to visualize and explore RDF data from SPARQL endpoints.

\spara{(4) ArangoDB} ({\scriptsize{https://www.arangodb.com/docs/stable/}}), which
is a document-based hybrid graph DB, provides a declarative query language AQL
(ArangoDB Query Language). It supports Apache TinkerPop Gremlin. ArangoDB has an
in-built graph viewer, additionally it supports third-party tools (e.g., Cytoscape)
for visualization and analysis. ArangoDB's graph ML tools provide several graph
embedding methods (e.g., GraphSage, Metapath2Vec, GAT, DMGI) over both homogeneous
and heterogeneous networks ({\scriptsize{https://github.com/arangoml/fastgraphml}}).

\spara{(5) OrientDB} ({\scriptsize{\url{https://orientdb.com}}}), a document-based native graph DB,
offers SQL extension for graph queries, and supports Gremlin. OrientDB studio
visualizes graphs and schema.

\spara{(6) JanusGraph} ({\scriptsize{\url{http://janusgraph.org}}})
uses a number of wide-column stores as backends, e.g.,
Apache Cassandra, HBase, Google Cloud Bigtable, Oracle BerkeleyDB,
ScyllaDB, etc. It supports Apache TinkerPop Gremlin. To visualize graphs stored in JanusGraph, one
can use third-party tools, e.g., Cytoscape, Gephi plugin for Apache TinkerPop, Graphexp
KeyLines by Cambridge Intelligence, Linkurious, etc.

\spara{(7) Amazon Neptune} ({\scriptsize{\url{https://aws.amazon.com/neptune}}})
is part of Amazon Web Services (AWS), supporting both RDF and property graph models, as well as Gremlin,
openCypher, and SPARQL query languages. The query results can be interactively visualized using Neptune Workbench.
Neptune uses GNN methods and the Deep Graph Library (DGL) to support
a number of graph ML tasks, including node and edge classification, regression, link prediction,
graph embedding (R-GCN), and KG embedding (TransE, DistMult, RotatE).

\spara{(8) GraphDB} ({\scriptsize{\url{https://www.ontotext.com/products/graphdb}}})
is an RDF database using SPARQL query language. It supports faceted search and third-party tools,
such as metaphactory, for interactive visualization.

\spara{(9) TigerGraph} \cite{Deutsch19}
is a native graph database with property graph data model and
GSQL language. TigerGraph GraphStudio provides a graphical interface
for interactive visualization and exploration.
TigerGraph's ML Workbench is a Jupyter-based Python development framework
that is inter-operable with popular deep learning frameworks such as
PyTorch Geometric, DGL, and supports graph embedding (Node2Vec,
Fast Random Projection, and Weisfeiler-Lehman).

\spara{(10) FaunaDB} ({\scriptsize{\url{https://fauna.com}}}) is a document-relational database
with property graph model and GraphQL API.

\spara{Summary.} The top-10 
commercial graph databases support various languages for querying of KGs -- as RDF triples or property graphs. Besides, many of them also provide interactive interfaces for visualization, querying, and exploration of KGs. Their support for in-built ML-based KG querying is limited. Only Amazon Neptune provides a few popular KG embedding methods such as TransE, DistMult, and RotatE. While many of these graph databases (e.g., AllegroGraph, ArangoDB, OrientDB) are multi-model, supporting multiple data models against a single backend, none of them has in-house system for storage and querying of multi-modal data, such as KGs with text, images, and multimedia.

\section{Future Directions}
\label{sec:future}

Knowledge graphs can support a holistic integration solution for multi-modal data arriving from heterogeneous sources. For instance, nodes and edges in a KG can have arbitrary number of properties of different types, e.g., tabular, key-value pairs, text, images, and multimedia.  Therefore, KGs can be a unified data model for complex data lake problems, to model cross-domain and diverse data. We conclude with a discussion about future work on KG querying.

\noindent $\bullet$ {\bf Vector data management and querying.} With the prevalence of KG embedding based query processing, managing and querying of vector data is critical. Data management community can contribute in this domain with high-dimensional data indexing, join, querying,
and geometric data processing.

\noindent $\bullet$ {\bf Scalable embedding learning.} Scaling knowledge graphs embedding 
is challenging {\cite{MohoneyWXRV21,LererWSLWBP19,ZhengSMTYDXZK20}}. The problem gets exacerbated when combined with
more complex data, such as KG+query embedding and multi-modal KG embedding.
Advanced techniques are required for scalable embedding learning of multi-modal KGs, e.g., with language models,
and conversational KG-QA with sequence-to-sequence models.

\noindent $\bullet$ {\bf Graph databases support for KG embedding.} Current graph DBMS support for ML-based KG querying is limited. In future, they can incorporate more KG embedding models, vector data management and query processing techniques, as well as enable multi-modal KG storage and query, more interactive means of KG querying such as NLQs and dialogues.

\noindent $\bullet$ {\bf Usability of KG querying methods.} Besides SPARQL, a number of KG querying approaches exist, e.g., query languages, keyword search, query-by-example, faceted search, visual query, natural language questions,
and conversational QA. It would be interesting to holistically compare them, understand their user-friendliness,
and categorize what is applicable in which domains.

\noindent $\bullet$ {\bf Suitability of KG embedding models.} A number of KG embedding models exist, such as translation-based models (TransE, TransD,
TransH) and semantic matching models (RESCAL, DistMult, ConvE). Different
models preserve various types of relation properties,
e.g., symmetry, antisymmetry, inversion, composition, complex mapping properties, etc. {\cite{SunDNT19}}.
One can analyze which properties are important for what queries, leading to a realization
of which KG embedding models are suitable for different KGs and queries.

\noindent $\bullet$ {\bf Explainability, interoperability, and multi-lingual KG querying.} There is an increasing focus on interpretability of deep learning methods over graph-structured data. In this context, explainability in knowledge graph embeddings is also important, for instance, what is being learned in knowledge graph embedding and KG-QA with explanatory evidences. Interoperability between KGs and supporting multi-lingual KGs {\cite{Pei0Y020}} and queries are other interesting future directions.

\section{Acknowledgement}
Arijit Khan acknowledges support from the Novo Nordisk Foundation grant NNF22OC0072415.

\balance

\begin{small}
\bibliographystyle{abbrv}
\bibliography{ref}

\begin{thebibliography}{100}

\bibitem{AbadiMMH09}
D.~J. Abadi, A.~Marcus, S.~Madden, and K.~Hollenbach.
\newblock Sw-store: a vertically partitioned {DBMS} for semantic web data
  management.
\newblock {\em {VLDB} J.}, 18(2):385--406, 2009.

\bibitem{AbdelazizHKK17}
I.~Abdelaziz, R.~Harbi, Z.~Khayyat, and P.~Kalnis.
\newblock A survey and experimental comparison of distributed {SPARQL} engines
  for very large {RDF} data.
\newblock {\em PVLDB}, 10(13):2049--2060, 2017.

\bibitem{Abu-Salih21}
B.~Abu{-}Salih.
\newblock Domain-specific knowledge graphs: a survey.
\newblock {\em J. Netw. Comput. Appl.}, 185:103076, 2021.

\bibitem{AbujabalRYW18}
A.~Abujabal, R.~S. Roy, M.~Yahya, and G.~Weikum.
\newblock Never-ending learning for open-domain question answering over
  knowledge bases.
\newblock In {\em WWW}, 2018.

\bibitem{AliBHVGSFTL22}
M.~Ali, M.~Berrendorf, C.~T. Hoyt, L.~Vermue, M.~Galkin, S.~Sharifzadeh,
  A.~Fischer, V.~Tresp, and J.~Lehmann.
\newblock Bringing light into the dark: a large-scale evaluation of knowledge
  graph embedding models under a unified framework.
\newblock {\em {IEEE} Trans. Pattern Anal. Mach. Intell.}, 44(12):8825--8845,
  2022.

\bibitem{AliSYHN22}
W.~Ali, M.~Saleem, B.~Yao, A.~Hogan, and A.~N. Ngomo.
\newblock A survey of {RDF} stores {\&} {SPARQL} engines for querying knowledge
  graphs.
\newblock {\em {VLDB} J.}, 31(3):1--26, 2022.

\bibitem{AnglesABBFGLPPS18}
R.~Angles, M.~Arenas, P.~Barcel{\'{o}}, P.~A. Boncz, G.~H.~L. Fletcher,
  C.~Gutierrez, T.~Lindaaker, M.~Paradies, S.~Plantikow, J.~F. Sequeda, O.~van
  Rest, and H.~Voigt.
\newblock {G-CORE:} a core for future graph query languages.
\newblock In {\em SIGMOD}, 2018.

\bibitem{AnglesABHRV17}
R.~Angles, M.~Arenas, P.~Barcel{\'{o}}, A.~Hogan, J.~L. Reutter, and D.~Vrgoc.
\newblock Foundations of modern query languages for graph databases.
\newblock {\em {ACM} Comput. Surv.}, 50(5):68:1--68:40, 2017.

\bibitem{ArakelyanDMC21}
E.~Arakelyan, D.~Daza, P.~Minervini, and M.~Cochez.
\newblock Complex query answering with neural link predictors.
\newblock In {\em ICLR}, 2021.

\bibitem{AGM17}
A.~Atserias, M.~Grohe, and D.~Marx.
\newblock Size bounds and query plans for relational joins.
\newblock {\em CoRR}, abs/1711.03860, 2017.

\bibitem{BalazevicAH19}
I.~Balazevic, C.~Allen, and T.~M. Hospedales.
\newblock Tucker: tensor factorization for knowledge graph completion.
\newblock In {\em EMNLP-IJCNLP}, 2019.

\bibitem{BalogK19}
K.~Balog and T.~Kenter.
\newblock Personal knowledge graphs: a research agenda.
\newblock In {\em SIGIR}, 2019.

\bibitem{BellomariniSG18}
L.~Bellomarini, E.~Sallinger, and G.~Gottlob.
\newblock The vadalog system: datalog-based reasoning for knowledge graphs.
\newblock {\em PVLDB}, 11(9):975--987, 2018.

\bibitem{BerantCFL13}
J.~Berant, A.~Chou, R.~Frostig, and P.~Liang.
\newblock Semantic parsing on freebase from question-answer pairs.
\newblock In {\em EMNLP}, 2013.

\bibitem{BPGFPBAH19}
M.~Besta, E.~Peter, R.~Gerstenberger, M.~Fischer, M.~Podstawski, C.~Barthels,
  G.~Alonso, and T.~Hoefler.
\newblock Demystifying graph databases: analysis and taxonomy of data
  organization, system designs, and graph queries.
\newblock {\em CoRR}, 2019.

\bibitem{BhowmickC22}
S.~S. Bhowmick and B.~Choi.
\newblock Data-driven visual query interfaces for graphs: past, present, and
  (near) future.
\newblock In {\em SIGMOD}, 2022.

\bibitem{BhowmickCL17}
S.~S. Bhowmick, B.~Choi, and C.~Li.
\newblock Graph querying meets {HCI:} state of the art and future directions.
\newblock In {\em SIGMOD}, 2017.

\bibitem{BollackerEPST08}
K.~D. Bollacker, C.~Evans, P.~K. Paritosh, T.~Sturge, and J.~Taylor.
\newblock Freebase: a collaboratively created graph database for structuring
  human knowledge.
\newblock In {\em SIGMOD}, 2008.

\bibitem{BonifatiCL15}
A.~Bonifati, R.~Ciucanu, and A.~Lemay.
\newblock Learning path queries on graph databases.
\newblock In {\em EDBT}, 2015.

\bibitem{2018Bonifati}
A.~Bonifati, G.~H.~L. Fletcher, H.~Voigt, and N.~Yakovets.
\newblock {\em Querying graphs}.
\newblock Synthesis Lectures on Data Management. Morgan {\&} Claypool
  Publishers, 2018.

\bibitem{BonifatiMT20}
A.~Bonifati, W.~Martens, and T.~Timm.
\newblock An analytical study of large {SPARQL} query logs.
\newblock {\em {VLDB} J.}, 29(2-3):655--679, 2020.

\bibitem{BordesUCW15}
A.~Bordes, N.~Usunier, S.~Chopra, and J.~Weston.
\newblock Large-scale simple question answering with memory networks.
\newblock {\em CoRR}, abs/1506.02075, 2015.

\bibitem{BordesUGWY13}
A.~Bordes, N.~Usunier, A.~Garc{\'{\i}}a{-}Dur{\'{a}}n, J.~Weston, and
  O.~Yakhnenko.
\newblock Translating embeddings for modeling multi-relational data.
\newblock In {\em NeurIPS}, 2013.

\bibitem{BorgwardtCL18}
S.~Borgwardt, {\.I}.~{\.I}. Ceylan, and T.~Lukasiewicz.
\newblock Recent advances in querying probabilistic knowledge bases.
\newblock In {\em IJCAI}, 2018.

\bibitem{BouhenniYNK21}
S.~Bouhenni, S.~Yahiaoui, N.~Nouali{-}Taboudjemat, and H.~Kheddouci.
\newblock A survey on distributed graph pattern matching in massive graphs.
\newblock {\em {ACM} Comput. Surv.}, 54(2):36:1--36:35, 2022.

\bibitem{CaiZWW21}
J.~Cai, Z.~Zhang, F.~Wu, and J.~Wang.
\newblock Deep cognitive reasoning network for multi-hop question answering
  over knowledge graphs.
\newblock In {\em ACL/IJCNLP}, 2021.

\bibitem{CMA12}
D.~V. Camarda, S.~Mazzini, and A.~Antonuccio.
\newblock Lodlive, exploring the web of data.
\newblock In {\em Semantic Systems}, 2012.

\bibitem{ChakrabortyLMTL21}
N.~Chakraborty, D.~Lukovnikov, G.~Maheshwari, P.~Trivedi, J.~Lehmann, and
  A.~Fischer.
\newblock Introduction to neural network-based question answering over
  knowledge graphs.
\newblock {\em WIREs Data Mining Knowl. Discov.}, 11(3), 2021.

\bibitem{ChenTCSZ18}
M.~Chen, Y.~Tian, K.~Chang, S.~Skiena, and C.~Zaniolo.
\newblock Co-training embeddings of knowledge graphs and entity descriptions
  for cross-lingual entity alignment.
\newblock In {\em IJCAI}, 2018.

\bibitem{ChenHS22}
X.~Chen, Z.~Hu, and Y.~Sun.
\newblock Fuzzy logic based logical query answering on knowledge graphs.
\newblock In {\em AAAI}, 2022.

\bibitem{ChongDES05}
E.~I. Chong, S.~Das, G.~Eadon, and J.~Srinivasan.
\newblock An efficient sql-based {RDF} querying scheme.
\newblock In {\em VLDB}, 2005.

\bibitem{ChristmannRASW19}
P.~Christmann, R.~S. Roy, A.~Abujabal, J.~Singh, and G.~Weikum.
\newblock Look before you hop: conversational question answering over knowledge
  graphs using judicious context expansion.
\newblock In {\em CIKM}, 2019.

\bibitem{DarariPN14}
F.~Darari, R.~E. Prasojo, and W.~Nutt.
\newblock {CORNER:} a completeness reasoner for {SPARQL} queries over {RDF}
  data sources.
\newblock In {\em The Semantic Web: {ESWC} Satellite Events}, 2014.

\bibitem{DettmersMS018}
T.~Dettmers, P.~Minervini, P.~Stenetorp, and S.~Riedel.
\newblock Convolutional 2d knowledge graph embeddings.
\newblock In {\em AAAI}, 2018.

\bibitem{DeutschFGHLLLMM22}
A.~Deutsch, N.~Francis, A.~Green, K.~Hare, B.~Li, L.~Libkin, T.~Lindaaker,
  V.~Marsault, W.~Martens, J.~Michels, F.~Murlak, S.~Plantikow, P.~Selmer,
  O.~van Rest, H.~Voigt, D.~Vrgoc, M.~Wu, and F.~Zemke.
\newblock Graph pattern matching in {GQL} and {SQL/PGQ}.
\newblock In {\em SIGMOD}, 2022.

\bibitem{DXW18}
A.~Deutsch, Y.~Xu, and M.~Wu.
\newblock Seamless syntactic and semantic integration of query primitives over
  relational and graph data in gsql, 2018.

\bibitem{Deutsch19}
A.~Deutsch, Y.~Xu, M.~Wu, and V.~E. Lee.
\newblock Tigergraph: a native {MPP} graph database.
\newblock {\em CoRR}, abs/1901.08248, 2019.

\bibitem{WangWGD18}
B.~Ding, Q.~Wang, B.~Wang, and L.~Guo.
\newblock Improving knowledge graph embedding using simple constraints.
\newblock In {\em ACL}, 2018.

\bibitem{DingYLHC022}
Y.~Ding, J.~Yu, B.~Liu, Y.~Hu, M.~Cui, and Q.~Wu.
\newblock Mukea: multimodal knowledge extraction and accumulation for
  knowledge-based visual question answering.
\newblock In {\em CVPR}, 2022.

\bibitem{DuLWCY22}
H.~Du, Z.~Le, H.~Wang, Y.~Chen, and J.~Yu.
\newblock {COKG-QA:} multi-hop question answering over {COVID-19} knowledge
  graphs.
\newblock {\em Data Intell.}, 4(3):471--492, 2022.

\bibitem{ElbedweihyWC12}
K.~Elbedweihy, S.~N. Wrigley, and F.~Ciravegna.
\newblock Evaluating semantic search query approaches with expert and casual
  users.
\newblock In {\em ISWC}, 2012.

\bibitem{FanLMTW11}
W.~Fan, J.~Li, S.~Ma, N.~Tang, and Y.~Wu.
\newblock Adding regular expressions to graph reachability and pattern queries.
\newblock In {\em ICDE}, 2011.

\bibitem{FanLMWW10}
W.~Fan, J.~Li, S.~Ma, H.~Wang, and Y.~Wu.
\newblock Graph homomorphism revisited for graph matching.
\newblock {\em PVLDB}, 3(1):1161--1172, 2010.

\bibitem{FanWW14}
W.~Fan, X.~Wang, and Y.~Wu.
\newblock Answering graph pattern queries using views.
\newblock In {\em ICDE}, 2014.

\bibitem{FZHWX22}
Q.~Fang, X.~Zhang, J.~Hu, X.~Wu, , and C.~Xu.
\newblock Contrastive multi-modal knowledge graph representation learning.
\newblock {\em {IEEE} Trans. Knowl. Data Eng.}, 2022.

\bibitem{FrancisGGLLMPRS18}
N.~Francis, A.~Green, P.~Guagliardo, L.~Libkin, T.~Lindaaker, V.~Marsault,
  S.~Plantikow, M.~Rydberg, P.~Selmer, and A.~Taylor.
\newblock Cypher: an evolving query language for property graphs.
\newblock In {\em SIGMOD}, 2018.

\bibitem{GZRT22}
M.~Galkin, Z.~Zhu, H.~Ren, and J.~Tang.
\newblock Inductive logical query answering in knowledge graphs.
\newblock In {\em NeurIPS}, 2022.

\bibitem{GareyJ79}
M.~R. Garey and D.~S. Johnson.
\newblock {\em Computers and intractability: a guide to the theory of
  NP-completeness}.
\newblock W. H. Freeman, 1979.

\bibitem{GeseseBAS21}
G.~A. Gesese, R.~Biswas, M.~Alam, and H.~Sack.
\newblock A survey on knowledge graph embeddings with literals: Which model
  links better literal-ly?
\newblock {\em Semantic Web}, 12(4):617--647, 2021.

\bibitem{GuoPH05}
Y.~Guo, Z.~Pan, and J.~Heflin.
\newblock Lubm: a benchmark for owl knowledge base systems.
\newblock {\em J. Web Semant.}, 3(2-3):158--182, 2005.

\bibitem{HamiltonBZJL18}
W.~L. Hamilton, P.~Bajaj, M.~Zitnik, D.~Jurafsky, and J.~Leskovec.
\newblock Embedding logical queries on knowledge graphs.
\newblock In {\em NeurIPS}, 2018.

\bibitem{HanFJ11}
L.~Han, T.~Finin, and A.~Joshi.
\newblock Gorelations: an intuitive query system for dbpedia.
\newblock In {\em JIST}, 2011.

\bibitem{HLPY10}
W.-S. Han, J.~Lee, M.-D. Pham, and J.~X. Yu.
\newblock {iGraph}: a framework for comparisons of disk-based graph indexing
  techniques.
\newblock {\em PVLDB}, 3(1–2):449–459, 2010.

\bibitem{HuX0WYZCS22}
Z.~Hu, Y.~Xu, W.~Yu, S.~Wang, Z.~Yang, C.~Zhu, K.~Chang, and Y.~Sun.
\newblock Empowering language models with knowledge graph reasoning for
  open-domain question answering.
\newblock In {\em EMNLP}, 2022.

\bibitem{HuangZLL19}
X.~Huang, J.~Zhang, D.~Li, and P.~Li.
\newblock Knowledge graph embedding based question answering.
\newblock In {\em WSDM}, 2019.

\bibitem{HuangCL22}
Z.~Huang, M.~Chiang, and W.~Lee.
\newblock Line: logical query reasoning over hierarchical knowledge graphs.
\newblock In {\em KDD}, 2022.

\bibitem{IbragimovHPZ16}
D.~Ibragimov, K.~Hose, T.~B. Pedersen, and E.~Zim{\'{a}}nyi.
\newblock Optimizing aggregate {SPARQL} queries using materialized {RDF} views.
\newblock In {\em ISWC}, 2016.

\bibitem{IlievskiSZ21}
F.~Ilievski, P.~A. Szekely, and B.~Zhang.
\newblock {CSKG:} the commonsense knowledge graph.
\newblock In {\em ESWC}, 2021.

\bibitem{ILL15}
M.~S. Islam, C.~Liu, and J.~Li.
\newblock Efficient answering of why-not questions in similar graph matching.
\newblock {\em {IEEE} Trans. Knowl. Data Eng.}, 27(10):2672--2686, 2015.

\bibitem{JayaramKLYE15}
N.~Jayaram, A.~Khan, C.~Li, X.~Yan, and R.~Elmasri.
\newblock Querying knowledge graphs by example entity tuples.
\newblock {\em {IEEE} Trans. Knowl. Data Eng.}, 27(10):2797--2811, 2015.

\bibitem{JiHXL015}
G.~Ji, S.~He, L.~Xu, K.~Liu, and J.~Zhao.
\newblock Knowledge graph embedding via dynamic mapping matrix.
\newblock In {\em ACL}, 2015.

\bibitem{JinY0YQP21}
X.~Jin, Z.~Yang, X.~Lin, S.~Yang, L.~Qin, and Y.~Peng.
\newblock {FAST:} fpga-based subgraph matching on massive graphs.
\newblock In {\em ICDE}, 2021.

\bibitem{KaloFEB20}
J.~Kalo, L.~Fichtel, P.~Ehler, and W.~Balke.
\newblock Knowlybert - hybrid query answering over language models and
  knowledge graphs.
\newblock In {\em ISWC}, 2020.

\bibitem{KaoudiM15}
Z.~Kaoudi and I.~Manolescu.
\newblock {RDF} in the clouds: a survey.
\newblock {\em {VLDB} J.}, 24(1):67--91, 2015.

\bibitem{KatsarouNT15}
F.~Katsarou, N.~Ntarmos, and P.~Triantafillou.
\newblock Performance and scalability of indexed subgraph query processing
  methods.
\newblock {\em Proc. {VLDB} Endow.}, 8(12):1566--1577, 2015.

\bibitem{KepuskaB18}
V.~Kepuska and G.~Bohouta.
\newblock Next-generation of virtual personal assistants (microsoft cortana,
  apple siri, amazon alexa and google home).
\newblock In {\em CCWC}, 2018.

\bibitem{KhanE14}
A.~Khan and S.~Elnikety.
\newblock Systems for big-graphs.
\newblock {\em Proc. {VLDB} Endow.}, 7(13):1709--1710, 2014.

\bibitem{KhanLYGCT11}
A.~Khan, N.~Li, X.~Yan, Z.~Guan, S.~Chakraborty, and S.~Tao.
\newblock Neighborhood based fast graph search in large networks.
\newblock In {\em SIGMOD}, 2011.

\bibitem{KhanSK18}
A.~Khan, G.~Segovia, and D.~Kossmann.
\newblock On smart query routing: for distributed graph querying with decoupled
  storage.
\newblock In {\em USENIX ATC}, 2018.

\bibitem{KhanWAY13}
A.~Khan, Y.~Wu, C.~C. Aggarwal, and X.~Yan.
\newblock Nema: fast graph search with label similarity.
\newblock {\em PVLDB}, 6(3):181--192, 2013.

\bibitem{2018Khan2}
A.~Khan, Y.~Ye, and L.~Chen.
\newblock {\em On uncertain graphs}.
\newblock Synthesis Lectures on Data Management. Morgan {\&} Claypool
  Publishers, 2018.

\bibitem{LeeHKL12}
J.~Lee, W.~Han, R.~Kasperovics, and J.~Lee.
\newblock An in-depth comparison of subgraph isomorphism algorithms in graph
  databases.
\newblock {\em Proc. {VLDB} Endow.}, 6(2):133--144, 2012.

\bibitem{LehmannIJJKMHMK15}
J.~Lehmann, R.~Isele, M.~Jakob, A.~Jentzsch, D.~Kontokostas, P.~N. Mendes,
  S.~Hellmann, M.~Morsey, P.~v.~Kleef, S.~Auer, and C.~Bizer.
\newblock Dbpedia - a large-scale, multilingual knowledge base extracted from
  wikipedia.
\newblock {\em Semantic Web}, 6(2):167--195, 2015.

\bibitem{LererWSLWBP19}
A.~Lerer, L.~Wu, J.~Shen, T.~Lacroix, L.~Wehrstedt, A.~Bose, and
  A.~Peysakhovich.
\newblock Pytorch-biggraph: a large scale graph embedding system.
\newblock In {\em MLSys}, 2019.

\bibitem{LiGC20}
Y.~Li, T.~Ge, and C.~X. Chen.
\newblock Online indices for predictive top-k entity and aggregate queries on
  knowledge graphs.
\newblock In {\em ICDE}, 2020.

\bibitem{LissandriniBV18}
M.~Lissandrini, M.~Brugnara, and Y.~Velegrakis.
\newblock Beyond macrobenchmarks: microbenchmark-based graph database
  evaluation.
\newblock {\em PVLDB}, 12(4):390--403, 2018.

\bibitem{LissandriniMPV20}
M.~Lissandrini, D.~Mottin, T.~Palpanas, and Y.~Velegrakis.
\newblock Graph-query suggestions for knowledge graph exploration.
\newblock In {\em WWW}, 2020.

\bibitem{LiuDJZT21}
L.~Liu, B.~Du, H.~Ji, C.~Zhai, and H.~Tong.
\newblock Neural-answering logical queries on knowledge graphs.
\newblock In {\em KDD}, 2021.

\bibitem{LiuDXXT22}
L.~Liu, B.~Du, J.~Xu, Y.~Xia, and H.~Tong.
\newblock Joint knowledge graph completion and question answering.
\newblock In {\em KDD}, 2022.

\bibitem{LiuZ0WJD020}
W.~Liu, P.~Zhou, Z.~Zhao, Z.~Wang, Q.~Ju, H.~Deng, and P.~Wang.
\newblock {K-BERT:} enabling language representation with knowledge graph.
\newblock In {\em AAAI}, 2020.

\bibitem{LiuZSCQZ0DT22}
X.~Liu, S.~Zhao, K.~Su, Y.~Cen, J.~Qiu, M.~Zhang, W.~Wu, Y.~Dong, and J.~Tang.
\newblock Mask and reason: pre-training knowledge graph transformers for
  complex logical queries.
\newblock In {\em KDD}, 2022.

\bibitem{LongZAWL22}
X.~Long, L.~Zhuang, L.~Aodi, S.~Wang, and H.~Li.
\newblock Neural-based mixture probabilistic query embedding for answering
  {FOL} queries on knowledge graphs.
\newblock In {\em EMNLP}, 2022.

\bibitem{MaLWCP19}
H.~Ma, M.~A. Langouri, Y.~Wu, F.~Chiang, and J.~Pi.
\newblock Ontology-based entity matching in attributed graphs.
\newblock {\em PVLDB}, 12(10):1195--1207, 2019.

\bibitem{MaCFHW14}
S.~Ma, Y.~Cao, W.~Fan, J.~Huai, and T.~Wo.
\newblock Strong simulation: capturing topology in graph pattern matching.
\newblock {\em {ACM} Trans. Database Syst.}, 39(1):4:1--4:46, 2014.

\bibitem{MhedhbiS22}
A.~Mhedhbi and S.~Salihoglu.
\newblock Modern techniques for querying graph-structured relations:
  foundations, system implementations, and open challenges.
\newblock {\em PVLDB}, 15(12):3762--3765, 2022.

\bibitem{microsoftsqlgraph}
Microsoft.
\newblock Sql graph architecture.
\newblock
  \url{https://learn.microsoft.com/en-us/sql/relational-databases/graphs/sql-graph-architecture?view=sql-server-ver16},
  2022.

\bibitem{MitchellCHTYBCM18}
T.~M. Mitchell, W.~W. Cohen, E.~R.~H. Jr., P.~P. Talukdar, B.~Yang,
  J.~Betteridge, A.~Carlson, B.~D. Mishra, M.~Gardner, B.~Kisiel,
  J.~Krishnamurthy, N.~Lao, K.~Mazaitis, T.~Mohamed, N.~Nakashole, E.~A.
  Platanios, A.~Ritter, M.~Samadi, B.~Settles, R.~C. Wang, D.~Wijaya, A.~Gupta,
  X.~Chen, A.~Saparov, M.~Greaves, and J.~Welling.
\newblock Never-ending learning.
\newblock {\em Commun. {ACM}}, 61(5):103--115, 2018.

\bibitem{MohoneyWXRV21}
J.~Mohoney, R.~Waleffe, H.~Xu, T.~Rekatsinas, and S.~Venkataraman.
\newblock Marius: learning massive graph embeddings on a single machine.
\newblock In {\em OSDI}, 2021.

\bibitem{MottinLVP14}
D.~Mottin, M.~Lissandrini, Y.~Velegrakis, and T.~Palpanas.
\newblock Exemplar queries: give me an example of what you need.
\newblock {\em PVLDB}, 7(5):365--376, 2014.

\bibitem{NathaniCSK19}
D.~Nathani, J.~Chauhan, C.~Sharma, and M.~Kaul.
\newblock Learning attention-based embeddings for relation prediction in
  knowledge graphs.
\newblock In {\em ACL}, 2019.

\bibitem{NavigliP10}
R.~Navigli and S.~P. Ponzetto.
\newblock Babelnet: building a very large multilingual semantic network.
\newblock In {\em ACL}, 2010.

\bibitem{neo4j}
Neo4J.
\newblock Why graph databases?
\newblock \url{https://neo4j.com/why-graph-databases/}, 2016.

\bibitem{NeumannW09}
T.~Neumann and G.~Weikum.
\newblock Scalable join processing on very large {RDF} graphs.
\newblock In {\em SIGMOD}, 2009.

\bibitem{NeumannW10}
T.~Neumann and G.~Weikum.
\newblock The {RDF-3X} engine for scalable management of {RDF} data.
\newblock {\em {VLDB} J.}, 19(1):91--113, 2010.

\bibitem{ngo18}
H.~Q. Ngo.
\newblock Worst-case optimal join algorithms: techniques, results, and open
  problems.
\newblock In {\em PODS}, 2018.

\bibitem{NguyenNNP18}
D.~Q. Nguyen, T.~D. Nguyen, D.~Q. Nguyen, and D.~Q. Phung.
\newblock A novel embedding model for knowledge base completion based on
  convolutional neural network.
\newblock In {\em NAACL-HLT}, 2018.

\bibitem{NiPYZC22}
J.~Ni, V.~Pandelea, T.~Young, H.~Zhou, and E.~Cambria.
\newblock Hitkg: towards goal-oriented conversations via multi-hierarchy
  learning.
\newblock In {\em AAAI}, 2022.

\bibitem{NickelTK11}
M.~Nickel, V.~Tresp, and H.~Kriegel.
\newblock A three-way model for collective learning on multi-relational data.
\newblock In {\em ICML}, 2011.

\bibitem{NickelTK12}
M.~Nickel, V.~Tresp, and H.~Kriegel.
\newblock Factorizing {YAGO:} scalable machine learning for linked data.
\newblock In {\em WWW}, 2012.

\bibitem{NikolaouK16}
C.~Nikolaou and M.~Koubarakis.
\newblock Querying incomplete information in {RDF} with {SPARQL}.
\newblock {\em Artif. Intell.}, 237:138--171, 2016.

\bibitem{NoyGJNPT19}
N.~F. Noy, Y.~Gao, A.~Jain, A.~Narayanan, A.~Patterson, and J.~Taylor.
\newblock Industry-scale knowledge graphs: lessons and challenges.
\newblock {\em Commun. {ACM}}, 62(8):36--43, 2019.

\bibitem{pgql}
Oracle.
\newblock Pgql 1.5 specification.
\newblock \url{https://pgql-lang.org/spec/1.5/}, 2022.

\bibitem{PacaciO19}
A.~Pacaci and M.~T. {\"{O}}zsu.
\newblock Experimental analysis of streaming algorithms for graph partitioning.
\newblock In {\em SIGMOD}, 2019.

\bibitem{PYHSH22}
X.~Pan, T.~Ye, D.~Han, S.~Song, and G.~Huang.
\newblock Contrastive language-image pre-training with knowledge graphs.
\newblock In {\em NeurIPS}, 2022.

\bibitem{PapailiouTKK15}
N.~Papailiou, D.~Tsoumakos, P.~Karras, and N.~Koziris.
\newblock Graph-aware, workload-adaptive {SPARQL} query caching.
\newblock In {\em SIGMOD}, 2015.

\bibitem{Pei0Y020}
S.~Pei, L.~Yu, G.~Yu, and X.~Zhang.
\newblock Rea: robust cross-lingual entity alignment between knowledge graphs.
\newblock In {\em KDD}, 2020.

\bibitem{PengGZOXZ21}
P.~Peng, Q.~Ge, L.~Zou, M.~T. {\"{O}}zsu, Z.~Xu, and D.~Zhao.
\newblock Optimizing multi-query evaluation in federated {RDF} systems.
\newblock {\em {IEEE} Trans. Knowl. Data Eng.}, 33(4):1692--1707, 2021.

\bibitem{PezeshkpourC018}
P.~Pezeshkpour, L.~Chen, and S.~Singh.
\newblock Embedding multimodal relational data for knowledge base completion.
\newblock In {\em EMNLP}, 2018.

\bibitem{PhuocQQNH16}
D.~L. Phuoc, H.~N.~M. Quoc, Q.~H. Ngo, T.~T. Nhat, and M.~Hauswirth.
\newblock The graph of things: a step towards the live knowledge graph of
  connected things.
\newblock {\em J. Web Semant.}, 37-38:25--35, 2016.

\bibitem{PoggiLCGLR08}
A.~Poggi, D.~Lembo, D.~Calvanese, G.~D. Giacomo, M.~Lenzerini, and R.~Rosati.
\newblock Linking data to ontologies.
\newblock {\em J. Data Semant.}, 10:133--173, 2008.

\bibitem{QuamarELO22}
A.~Quamar, V.~Efthymiou, C.~Lei, and F.~{\"{O}}zcan.
\newblock Natural language interfaces to data.
\newblock {\em Found. Trends Databases}, 11(4):319--414, 2022.

\bibitem{RahmB01}
E.~Rahm and P.~A. Bernstein.
\newblock A survey of approaches to automatic schema matching.
\newblock {\em {VLDB} J.}, 10(4):334--350, 2001.

\bibitem{RenDDCYSSLZ21}
H.~Ren, H.~Dai, B.~Dai, X.~Chen, M.~Yasunaga, H.~Sun, D.~Schuurmans,
  J.~Leskovec, and D.~Zhou.
\newblock {LEGO:} latent execution-guided reasoning for multi-hop question
  answering on knowledge graphs.
\newblock In {\em ICML}, 2021.

\bibitem{RenDDCZLS22}
H.~Ren, H.~Dai, B.~Dai, X.~Chen, D.~Zhou, J.~Leskovec, and D.~Schuurmans.
\newblock {SMORE:} knowledge graph completion and multi-hop reasoning in
  massive knowledge graphs.
\newblock In {\em KDD}, 2022.

\bibitem{RenHL20}
H.~Ren, W.~Hu, and J.~Leskovec.
\newblock Query2box: reasoning over knowledge graphs in vector space using box
  embeddings.
\newblock In {\em ICLR}, 2020.

\bibitem{RenL20}
H.~Ren and J.~Leskovec.
\newblock Beta embeddings for multi-hop logical reasoning in knowledge graphs.
\newblock In {\em NeurIPS}, 2020.

\bibitem{RenW16}
X.~Ren and J.~Wang.
\newblock Multi-query optimization for subgraph isomorphism search.
\newblock {\em PVLDB}, 10(3):121--132, 2016.

\bibitem{Rodriguez15}
M.~A. Rodriguez.
\newblock The gremlin graph traversal machine and language (invited talk).
\newblock In {\em DBPL}, 2015.

\bibitem{RoyA20}
R.~S. Roy and A.~Anand.
\newblock Question answering over curated and open web sources.
\newblock In {\em SIGIR}, 2020.

\bibitem{SagiLPH22}
T.~Sagi, M.~Lissandrini, T.~B. Pedersen, and K.~Hose.
\newblock A design space for {RDF} data representations.
\newblock {\em {VLDB} J.}, 31(2):347--373, 2022.

\bibitem{SahuMSLO20}
S.~Sahu, A.~Mhedhbi, S.~Salihoglu, J.~Lin, and M.~T. {\"{O}}zsu.
\newblock The ubiquity of large graphs and surprising challenges of graph
  processing: extended survey.
\newblock {\em {VLDB} J.}, 29(2-3):595--618, 2020.

\bibitem{SakrEH12}
S.~Sakr, S.~Elnikety, and Y.~He.
\newblock {G-SPARQL:} a hybrid engine for querying large attributed graphs.
\newblock In {\em CIKM}, 2012.

\bibitem{SarwatEHM13}
M.~Sarwat, S.~Elnikety, Y.~He, and M.~F. Mokbel.
\newblock Horton+: a distributed system for processing declarative reachability
  queries over partitioned graphs.
\newblock {\em PVLDB}, 6(14):1918--1929, 2013.

\bibitem{SaxenaKG22}
A.~Saxena, A.~Kochsiek, and R.~Gemulla.
\newblock Sequence-to-sequence knowledge graph completion and question
  answering.
\newblock In {\em ACL}, 2022.

\bibitem{SaxenaTT20}
A.~Saxena, A.~Tripathi, and P.~P. Talukdar.
\newblock Improving multi-hop question answering over knowledge graphs using
  knowledge base embeddings.
\newblock In {\em ACL}, 2020.

\bibitem{SchlichtkrullKB18}
M.~S. Schlichtkrull, T.~N. Kipf, P.~Bloem, R.~v.~d. Berg, I.~Titov, and
  M.~Welling.
\newblock Modeling relational data with graph convolutional networks.
\newblock In {\em ESWC}, 2018.

\bibitem{SHLP09}
M.~Schmidt, T.~Hornung, G.~Lausen, and C.~Pinkel.
\newblock Sp2bench: a sparql performance benchmark.
\newblock In {\em ICDE}, 2009.

\bibitem{ShiLZSY17}
C.~Shi, Y.~Li, J.~Zhang, Y.~Sun, and P.~S. Yu.
\newblock A survey of heterogeneous information network analysis.
\newblock {\em {IEEE} Trans. Knowl. Data Eng.}, 29(1):17--37, 2017.

\bibitem{google}
A.~Singhal.
\newblock Introducing the knowledge graph: things, not strings.
\newblock
  \url{https://blog.google/products/search/introducing-knowledge-graph-things-not/},
  2012.

\bibitem{Sommer14}
C.~Sommer.
\newblock Shortest-path queries in static networks.
\newblock {\em {ACM} Comput. Surv.}, 46(4):45:1--45:31, 2014.

\bibitem{SongNLW22}
Q.~Song, M.~H. Namaki, P.~Lin, and Y.~Wu.
\newblock Answering why-questions for subgraph queries.
\newblock {\em {IEEE} Trans. Knowl. Data Eng.}, 34(10):4636--4649, 2022.

\bibitem{speer-havasi-2012-representing}
R.~Speer and C.~Havasi.
\newblock Representing general relational knowledge in {C}oncept{N}et 5.
\newblock In {\em LREC}, 2012.

\bibitem{StyperekCSM15}
A.~Styperek, M.~Ciesielczyk, A.~Szwabe, and P.~Misiorek.
\newblock Evaluation of sparql-compliant semantic search user interfaces.
\newblock {\em Vietnam. J. Comput. Sci.}, 2(3):191--199, 2015.

\bibitem{SuYSSKVY15}
Y.~Su, S.~Yang, H.~Sun, M.~Srivatsa, S.~Kase, M.~Vanni, and X.~Yan.
\newblock Exploiting relevance feedback in knowledge graph search.
\newblock In {\em KDD}, 2015.

\bibitem{SKW07}
F.~M. Suchanek, G.~Kasneci, and G.~Weikum.
\newblock Yago: a core of semantic knowledge.
\newblock In {\em WWW}, 2007.

\bibitem{SunAB0C20}
H.~Sun, A.~O. Arnold, T.~Bedrax{-}Weiss, F.~Pereira, and W.~W. Cohen.
\newblock Faithful embeddings for knowledge base queries.
\newblock In {\em NeurIPS}, 2020.

\bibitem{SunCZWZZWZ20}
R.~Sun, X.~Cao, Y.~Zhao, J.~Wan, K.~Zhou, F.~Zhang, Z.~Wang, and K.~Zheng.
\newblock Multi-modal knowledge graphs for recommender systems.
\newblock In {\em CIKM}, 2020.

\bibitem{SL20}
S.~Sun and Q.~Luo.
\newblock In-memory subgraph matching: an in-depth study.
\newblock In {\em SIGMOD}, page 1083–1098, 2020.

\bibitem{SunFSKHX15}
W.~Sun, A.~Fokoue, K.~Srinivas, A.~Kementsietsidis, G.~Hu, and G.~T. Xie.
\newblock Sqlgraph: an efficient relational-based property graph store.
\newblock In {\em SIGMOD}, 2015.

\bibitem{SunDNT19}
Z.~Sun, Z.~Deng, J.~Nie, and J.~Tang.
\newblock Rotate: knowledge graph embedding by relational rotation in complex
  space.
\newblock In {\em ICLR}, 2019.

\bibitem{SzarnyasPAMPKEB18}
G.~Sz{\'{a}}rnyas, A.~Prat{-}P{\'{e}}rez, A.~Averbuch, J.~Marton, M.~Paradies,
  M.~Kaufmann, O.~Erling, P.~A. Boncz, V.~Haprian, and J.~B. Antal.
\newblock An early look at the {LDBC} social network benchmark's business
  intelligence workload.
\newblock In {\em GRADES and NDA}, 2018.

\bibitem{TalmorB18}
A.~Talmor and J.~Berant.
\newblock The web as a knowledge-base for answering complex questions.
\newblock In {\em NAACL-HLT}, 2018.

\bibitem{TchechmedjievFB19}
A.~Tchechmedjiev, P.~Fafalios, K.~Boland, M.~Gasquet, M.~Zloch, B.~Zapilko,
  S.~Dietze, and K.~Todorov.
\newblock Claimskg: a knowledge graph of fact-checked claims.
\newblock In {\em ISWC}, 2019.

\bibitem{TermehchyWCG12}
A.~Termehchy, M.~Winslett, Y.~Chodpathumwan, and A.~Gibbons.
\newblock Design independent query interfaces.
\newblock {\em {IEEE} Trans. Knowl. Data Eng.}, 24(10):1819--1832, 2012.

\bibitem{T22}
Y.~Tian.
\newblock The world of graph databases from an industry perspective.
\newblock {\em {SIGMOD} Rec.}, 51(4):60--67, 2022.

\bibitem{trivedi2017lc}
P.~Trivedi, G.~Maheshwari, M.~Dubey, and J.~Lehmann.
\newblock Lc-quad: a corpus for complex question answering over knowledge
  graphs.
\newblock In {\em ISWC}, 2017.

\bibitem{TuanCL19}
Y.~Tuan, Y.~Chen, and H.~Lee.
\newblock Dykgchat: benchmarking dialogue generation grounding on dynamic
  knowledge graphs.
\newblock In {\em EMNLP-IJCNLP}, 2019.

\bibitem{TzitzikasMP17}
Y.~Tzitzikas, N.~Manolis, and P.~Papadakos.
\newblock Faceted exploration of {RDF/S} datasets: a survey.
\newblock {\em J. Intell. Inf. Syst.}, 48(2):329--364, 2017.

\bibitem{VashishthSNT20}
S.~Vashishth, S.~Sanyal, V.~Nitin, and P.~P. Talukdar.
\newblock Composition-based multi-relational graph convolutional networks.
\newblock In {\em ICLR}, 2020.

\bibitem{VasilyevaTBL16}
E.~Vasilyeva, M.~Thiele, C.~Bornh{\"{o}}vd, and W.~Lehner.
\newblock Answering "why empty?" and "why so many?" queries in graph databases.
\newblock {\em J. Comput. Syst. Sci.}, 82(1):3--22, 2016.

\bibitem{VK14}
D.~Vrande\v{c}i\'{c} and M.~Kr\"{o}tzsch.
\newblock Wikidata: a free collaborative knowledgebase.
\newblock {\em Commun. ACM}, 57(10):78–85, 2014.

\bibitem{WZWL22}
L.~Wang, W.~Zhao, Z.~Wei, and J.~Liu.
\newblock Simkgc: simple contrastive knowledge graph completion with
  pre-trained language models.
\newblock In {\em ACL}, 2022.

\bibitem{WangWYZCQ21}
M.~Wang, S.~Wang, H.~Yang, Z.~Zhang, X.~Chen, and G.~Qi.
\newblock Is visual context really helpful for knowledge graph? a
  representation learning perspective.
\newblock In {\em MM}, 2021.

\bibitem{WangGZZLLT21}
X.~Wang, T.~Gao, Z.~Zhu, Z.~Zhang, Z.~Liu, J.~Li, and J.~Tang.
\newblock Kepler: a unified model for knowledge embedding and pre-trained
  language representation.
\newblock {\em Trans. Assoc. Comput. Linguistics}, 9:176--194, 2021.

\bibitem{WKWJY20}
Y.~Wang, A.~Khan, T.~Wu, J.~Jin, and H.~Yan.
\newblock Semantic guided and response times bounded top-k similarity search
  over knowledge graphs.
\newblock In {\em ICDE}, 2020.

\bibitem{WangKXJHF22}
Y.~Wang, A.~Khan, X.~Xu, J.~Jin, Q.~Hong, and T.~Fu.
\newblock Aggregate queries on knowledge graphs: fast approximation with
  semantic-aware sampling.
\newblock In {\em ICDE}, 2022.

\bibitem{WangKXYPZ22}
Y.~Wang, A.~Khan, X.~Xu, S.~Ye, S.~Pan, and Y.~Zhou.
\newblock Approximate and interactive processing of aggregate queries on
  knowledge graphs: a demonstration.
\newblock In {\em CIKM}, 2022.

\bibitem{WangLFYC21}
Y.~Wang, Y.~Li, J.~Fan, C.~Ye, and M.~Chai.
\newblock A survey of typical attributed graph queries.
\newblock {\em World Wide Web}, 24(1):297--346, 2021.

\bibitem{WangLLZ19}
Z.~Wang, L.~Li, Q.~Li, and D.~Zeng.
\newblock Multimodal data enhanced representation learning for knowledge
  graphs.
\newblock In {\em IJCNN}, 2019.

\bibitem{WangZFC14}
Z.~Wang, J.~Zhang, J.~Feng, and Z.~Chen.
\newblock Knowledge graph embedding by translating on hyperplanes.
\newblock In {\em AAAI}, 2014.

\bibitem{WeikumDRS21}
G.~Weikum, X.~L. Dong, S.~Razniewski, and F.~M. Suchanek.
\newblock Machine knowledge: creation and curation of comprehensive knowledge
  bases.
\newblock {\em Found. Trends Databases}, 10(2-4):108--490, 2021.

\bibitem{WilkinsonSKR03}
K.~Wilkinson, C.~Sayers, H.~A. Kuno, and D.~Reynolds.
\newblock Efficient {RDF} storage and retrieval in jena2.
\newblock In {\em SWDB}, 2003.

\bibitem{Wood12}
P.~T. Wood.
\newblock Query languages for graph databases.
\newblock {\em {SIGMOD} Rec.}, 41(1):50--60, 2012.

\bibitem{WHCS18}
M.~Wylot, M.~Hauswirth, P.~Cudr\'{e}-Mauroux, and S.~Sakr.
\newblock Rdf data storage and query processing schemes: a survey.
\newblock {\em ACM Comput. Surv.}, 51(4), 2018.

\bibitem{XieLJLS16}
R.~Xie, Z.~Liu, J.~Jia, H.~Luan, and M.~Sun.
\newblock Representation learning of knowledge graphs with entity descriptions.
\newblock In {\em AAAI}, 2016.

\bibitem{XieLLS17}
R.~Xie, Z.~Liu, H.~Luan, and M.~Sun.
\newblock Image-embodied knowledge representation learning.
\newblock In {\em IJCAI}, 2017.

\bibitem{XuRKKA20}
D.~Xu, C.~Ruan, E.~K{\"{o}}rpeoglu, S.~Kumar, and K.~Achan.
\newblock Product knowledge graph embedding for e-commerce.
\newblock In {\em WSDM}, 2020.

\bibitem{YangYHGD14a}
B.~Yang, W.~Yih, X.~He, J.~Gao, and L.~Deng.
\newblock Embedding entities and relations for learning and inference in
  knowledge bases.
\newblock In {\em ICLR}, 2015.

\bibitem{YangQLLL22}
D.~Yang, P.~Qing, Y.~Li, H.~Lu, and X.~Lin.
\newblock Gammae: gamma embeddings for logical queries on knowledge graphs.
\newblock In {\em EMNLP}, 2022.

\bibitem{YangYZ21}
J.~Yang, W.~Yao, and W.~Zhang.
\newblock Keyword search on large graphs: a survey.
\newblock {\em Data Sci. Eng.}, 6(2):142--162, 2021.

\bibitem{YangWSY14}
S.~Yang, Y.~Wu, H.~Sun, and X.~Yan.
\newblock Schemaless and structureless graph querying.
\newblock {\em PVLDB}, 7(7):565--576, 2014.

\bibitem{Yang0ZBCSM18}
Z.~Yang, P.~Qi, S.~Zhang, Y.~Bengio, W.~W. Cohen, R.~Salakhutdinov, and C.~D.
  Manning.
\newblock Hotpotqa: a dataset for diverse, explainable multi-hop question
  answering.
\newblock In {\em EMNLP}, 2018.

\bibitem{YaoCHH12}
J.~Yao, B.~Cui, L.~Hua, and Y.~Huang.
\newblock Keyword query reformulation on structured data.
\newblock In {\em ICDE}, 2012.

\bibitem{YBRZMLL22}
M.~Yasunaga, A.~Bosselut, H.~Ren, X.~Zhang, C.~D. Manning, P.~Liang, and
  J.~Leskovec.
\newblock Deep bidirectional language-knowledge graph pretraining.
\newblock In {\em NeurIPS}, 2022.

\bibitem{Yu0Y022}
D.~Yu, C.~Zhu, Y.~Yang, and M.~Zeng.
\newblock Jaket: joint pre-training of knowledge graph and language
  understanding.
\newblock In {\em AAAI}, 2022.

\bibitem{ZaibZSMZ22}
M.~Zaib, W.~E. Zhang, Q.~Z. Sheng, A.~Mahmood, and Y.~Zhang.
\newblock Conversational question answering: a survey.
\newblock {\em Knowl. Inf. Syst.}, 64(12):3151--3195, 2022.

\bibitem{ZengZOHZ20}
L.~Zeng, L.~Zou, M.~T. {\"{O}}zsu, L.~Hu, and F.~Zhang.
\newblock {GSI:} gpu-friendly subgraph isomorphism.
\newblock In {\em ICDE}, 2020.

\bibitem{ZhangD0H22}
M.~Zhang, R.~Dai, M.~Dong, and T.~He.
\newblock Drlk: dynamic hierarchical reasoning with language model and
  knowledge graph for question answering.
\newblock In {\em EMNLP}, 2022.

\bibitem{abs-2202-07412}
W.~Zhang, J.~Chen, J.~Li, Z.~Xu, J.~Z. Pan, and H.~Chen.
\newblock Knowledge graph reasoning with logics and embeddings: survey and
  perspective.
\newblock {\em CoRR}, abs/2202.07412, 2022.

\bibitem{ZhangPWCZZBC19}
W.~Zhang, B.~Paudel, L.~Wang, J.~Chen, H.~Zhu, W.~Zhang, A.~Bernstein, and
  H.~Chen.
\newblock Iteratively learning embeddings and rules for knowledge graph
  reasoning.
\newblock In {\em WWW}, 2019.

\bibitem{ZhangCTW13}
X.~Zhang, L.~Chen, Y.~Tong, and M.~Wang.
\newblock {EAGRE:} towards scalable {I/O} efficient {SPARQL} query evaluation
  on the cloud.
\newblock In {\em ICDE}, 2013.

\bibitem{ZhangWCJW21}
Z.~Zhang, J.~Wang, J.~Chen, S.~Ji, and F.~Wu.
\newblock Cone: cone embeddings for multi-hop reasoning over knowledge graphs.
\newblock In {\em NeurIPS}, 2021.

\bibitem{ZhengSMTYDXZK20}
D.~Zheng, X.~Song, C.~Ma, Z.~Tan, Z.~Ye, J.~Dong, H.~Xiong, Z.~Zhang, and
  G.~Karypis.
\newblock Dgl-ke: training knowledge graph embeddings at scale.
\newblock In {\em SIGIR}, 2020.

\bibitem{ZhengYZC18}
W.~Zheng, J.~X. Yu, L.~Zou, and H.~Cheng.
\newblock Question answering over knowledge graphs: question understanding via
  template decomposition.
\newblock {\em Proc. {VLDB} Endow.}, 11(11):1373--1386, 2018.

\bibitem{ZhengZPYSZ16}
W.~Zheng, L.~Zou, W.~Peng, X.~Yan, S.~Song, and D.~Zhao.
\newblock Semantic {SPARQL} similarity search over {RDF} knowledge graphs.
\newblock {\em PVLDB}, 9(11):840--851, 2016.

\bibitem{ZhouYHZXZ18}
H.~Zhou, T.~Young, M.~Huang, H.~Zhao, J.~Xu, and X.~Zhu.
\newblock Commonsense knowledge aware conversation generation with graph
  attention.
\newblock In {\em IJCAI}, 2018.

\bibitem{ZhouWXWY07}
Q.~Zhou, C.~Wang, M.~Xiong, H.~Wang, and Y.~Yu.
\newblock {SPARK:} adapting keyword query to semantic search.
\newblock In {\em ISWC}. Springer, 2007.

\bibitem{Zhu0Z022}
Z.~Zhu, M.~Galkin, Z.~Zhang, and J.~Tang.
\newblock Neural-symbolic models for logical queries on knowledge graphs.
\newblock In {\em ICML}, 2022.

\bibitem{ZouHWYHZ14}
L.~Zou, R.~Huang, H.~Wang, J.~X. Yu, W.~He, and D.~Zhao.
\newblock Natural language question answering over {RDF:} a graph data driven
  approach.
\newblock In {\em SIGMOD}, 2014.

\end{thebibliography}
\end{small}

\end{document}